\documentclass[12pt]{article}

\usepackage{gthstyle}
 \usepackage{amsfonts}
 \usepackage{amssymb,amsmath}
 \usepackage{graphicx} \usepackage{epstopdf}
 \newcommand{\crlb}[1]{\label{#1}\\[2pt]}
 
 \newcommand{\crld}[1]{\label{#1}}
 \newcommand{\eela}[1]{\quad\hbox{\scriptsize{#1}}\label{#1}\end{eqnarray}}
 \newcommand{\eelb}[1]{\label{#1}\end{eqnarray}}
 
 \newcommand{\newsecb}[2]{\section{#1}\label{#2}\setcounter{equation}{0}}

 \newcommand{\nolabels} {\def\eel{\eelb} \def\crl{\crlb} \def\newsecl{\newsecb}\def\bibiteml{\bibitem}\def\citel{\cite}\def\labell{\crld}}

 \newcommand\publishversion{\nolabels\setlength{\textheight}{8.35in}\setlength{\oddsidemargin}{0in}
    \setlength{\textwidth}{6.3in}\setlength{\topmargin}{-0.3in}}

                 \def\fn{\footnote}
        \def\be{\begin{eqnarray}}    \def\ee{\end{eqnarray}}
 \def\bi#1{\begin{itemize}\item[#1]}   \def\itm#1{\item[#1]}  \def\ei{\end{itemize}}  \def\eqn#1{(\ref{#1})}
 \def\tl#1{\tilde{#1}}  \def\^#1{\hat{#1}}

       \def\b{\beta}         
 \def\d{\delta}      \def\D{\Delta}  \def\e{\varepsilon} 
             
         \def\F{\Phi}    \def\vv{\varphi}    
             \def\r{\varrho}       
         \def\tht{\theta}  
               
       \def\W{\Omega}  

 \def\HH{{\mathcal H}}        \def\MM{{\mathcal M}}
  \def\ra{\rightarrow}
  
 \def\dd{{\rm d}}     \def\ket{\rangle}

 \def\iss{\ =\ }

 \def\fract#1#2{{\textstyle{#1\over#2}}}
 \def\ffract#1#2{\raise .2 em\hbox{$\scriptstyle#1$}\kern-.3em/
                 \kern-.2em\lower .15 em \hbox{$\scriptstyle#2$}}
 
 \def\half{\fract12}  

 \def\ex#1{e^{\textstyle#1}} \def\qqquad{\qquad\qquad}
 \def\bpm{\begin{pmatrix}} \def\epm{\end{pmatrix}}
  
\def\bmatrix{\begin{matrix}} \def\ematrix{\end{matrix}} \def\bpmatrix{\begin{pmatrix}}\def\epmatrix{\end{pmatrix}}
\def\bcenter{\begin{center}} \def\ecenter{\end{center}}

\def\lowerheightfig#1#2#3{\(\raise-#1\hbox{\includegraphics[height=#2]{#3}}\)}
\def\lowerwidthfig#1#2#3{\(\raise-#1\hbox{\includegraphics[width=#2]{#3}}\)}
\def\inn{{\mathrm{in}}}  \def\outt{{\mathrm{out}}}

  \publishversion
 
\begin{document}

\bcenter 
{ \LARGE\textbf{The Quantum Black Hole as a Hydrogen Atom: Microstates Without Strings Attached \\[20pt] }}
{\large Gerard 't~Hooft}  \\[15pt]
Institute for Theoretical Physics \\[5pt]
\(\mathrm{EMME}\F\)   \\
Centre for Extreme Matter and Emergent Phenomena\\[5pt] 
Science Faculty,  Utrecht University\\[5pt]
 POBox 80.195,
 3808TD\ Utrecht  \\
The Netherlands  \\[10pt] 
e-mail:  g.thooft@uu.nl 
\\ internet:  http://www.staff.science.uu.nl/\~{}hooft101/ 
\ecenter
\vfil
\begin{quotation} \noindent {\large\bf Abstract } \\

{\small Applying an expansion in spherical harmonics, turns the black hole with its microstates into something about as transparent as the hydrogen atom was in the early days of quantum mechanics. It enables us to present a concise description of the evolution laws of these microstates, linking them to perturbative quantum field theory, in the background of the Schwarzschild metric. Three pieces of insight are obtained:
One, we learn how the gravitational back reaction, whose dominant component can be calculated exactly, turns particles entering the hole, into particles leaving it, by exchanging the momentum- and position operators; two, we find out how this effect removes firewalls, both on the future and the past event horizon, and three,  we discover that the presence of region $II$ in the Penrose diagram forces a topological twist in the background metric, culminating in antipodal identification. Although a cut-off is required that effectively replaces the transverse coordinates by a lattice,  the effect of such a cut-off  minimizes when the spherical wave expansion is applied. This expansion then reveals exactly how antipodal identification restores unitarity -- for each partial wave separately. We expect that these ideas will provide new insights in some highly non-trivial topological space-time features at the Planck scale.}
\end{quotation}
\vfil 
\noindent Version: May 2016 \   \\[2pt]
{\footnotesize Last typeset: \today}
\eject
\setcounter{page}{2} 

\newsecl{Introduction}{intro}

A line of research has been set up by the author, with the aim of obtaining further insights in the texture of space, time and matter, by making one simple basic assumption: whatever happens during gravitational collapse, should be naturally incorporated in our theories of particles and forces. If we try to describe something like black holes, their behavior should be understood in the same language as the one we use for other particles; black holes should be treated just like atoms, molecules and nano particles, apart perhaps from being just a bit more exotic. If we find it useful to describe particles in terms of quantum wave equations, virtual particle states, and other such notions, we should  expect that these should also apply to black holes;  there will be virtual black holes, and so on.

The danger of this approach in asking our questions can be recognized in much of the present literature: prejudices and assumptions. Many of these are much less likely to generate the right theories than usually thought. And so, we decided to move forwards with steps that are as small as possible, so as to avoid mistakes. We are not afraid of making mistakes, but we do wish to spot all weaknesses that we will encounter, and my taste in these matters seems to deviate from many others, notably when people talk of string theories, extra dimensions and what not. 

In fundamental physics, there is the important question of strategy. 
Strings may be right, extra dimensions may exist, but if this is so, we wish to derive those from first principles. We suspect that most theoreticians today, asking basic questions about Nature, are too ambitious in searching for `unification' and the like\,\cite{smolin}. They make too many wild guesses. In our view, there is ample information in what we already know, but not fully understand, even when it sounds as boring as perturbative quantum field theory and perturbative gravity in the Schwarzschild background. Thus, we wish to use well-established knowledge and just inch our way forward.  What we observe is, that such an approach is possible, and extremely rewarding, but it slows us down a lot. 

In short, our present approach does not come with wild sceneries, new math languages, wild speculations, but only with small tangible facts. We do produce new results this way, notably about the topology in space-time and its non-commutative nature. Both are not what most people think; they are more interesting than that.

Black holes must be quantum forms of matter, moving in accordance with Schr\"odinger equations just like everything else. At the same time, we have General Relativity. An observer moving close to, or perhaps into, a black hole, must experience laws of physics that are in accordance with the tidal forces that (s)he should expect to feel. This leads to the first honest calculation one can do: apply the laws of quantum field theory near the black hole horizon. The calculation was done by Hawking\,\cite{SWH} with the well-known result that a distant observer will see particles emitted by the black hole. However, it seems that such black holes would then sport a kind of behaviour that is not in line with our expectations: they radiate strictly thermally, without any Schr\"odinger equation applying to the black holes as objects. This cannot be right.

A correct description of the quantum states of a massive object such as a black hole should comprise exactly all quantum states it can be in, which form a basis in Hilbert space. These states interact with matter falling in, and determine the dynamics of the states of particles coming out. These quantum laws should not depend strongly on how this object was formed millions of years ago, or what it plans to become millions of years from now.

Take a Kerr-Newman black hole at a given energy, corresponding to its mass \(\MM\equiv GM\), where \(G\) is Newton's constant\fn{We keep Newton's constant because we will be interested to know how our results depend on it, keeping the rest fixed.}, while \(c=\hbar=1\). Its entropy is given by the horizon area \(A\), known to be given by

	\be \frac A{4\pi}=2\MM^2-Q^2+2\MM\sqrt{\MM^2-Q^2-(J/\MM)^2}\ , \eel{genentropy}
for a Kerr-Newman black hole with mass \(M\), charge \(Q\), and angular momentum \(J\) in natural units.

We see that, when we keep \(\MM\) fixed, the entropy maximizes when \(Q=J=0\), so that, at given energy, by far the largest number of quantum states occur at \(Q\approx J\approx 0\), the pure Schwarzschild black hole. For the generic quantum state, \(Q\) and \(J\) must average out to zero, and this justifies that we ignore \(Q\) and \(J\), for the time being. They can be treated perturbatively later.

Furthermore, the quantum world as we know it, is symmetric under time reversal (more precisely, \(CPT\) reversal). Therefore, as long as we rely on known laws of physics, our description of the generic quantum state of a black hole should not be based on Penrose diagrams that are not symmetric under time reversal.

On the other hand, the gravitational force acting between in-going and out-going matter cannot be ignored, as we observed in earlier work\,\cite{GtHBH85,GtHgravshift,GtHSmatrix}. Thus,  we think the generic quantum state of a black hole can be studied in a way that is fundamentally better than what we usually see in the literature, by taking gravitational back reaction into account. It just so happens that the most important ingredient of the gravitational force needed here, is described by \emph{linear equations}, and this enables us to disentangle the effects, with quite revealing results.

\newsecl{A reformulation of previous results}{previous}
\subsection{The classical theory\labell{classical}}
We refer to our previous paper on this subject\,\cite{GtHantipodal} for the technical explanations of the calculations that will corroborate the scenario explained later in this chapter and in subsequent ones. This entire chapter mainly consists of older material.

Our present aim is to precisely formulate the behaviour of the black hole microstates. First, consider the classical (i.e., non quantum mechanical) problem. It consists of a straightforward application of General Relativity without anything else added. We have the Schwarzschild metric,
	\be \dd s^2=-\left(1-\frac{2\mathcal{M}}r\right)\,\dd t^2+\frac{\dd r^2}{1-2\mathcal{M}/r}+r^2\,\dd\W^2\ , \eel{SSmetric}
where \(\dd\W^2\) stands short for the angular part of the metric, \(\dd\W^2\equiv\dd\tht^2+\sin^2\tht\,\dd\vv^2\), and \(\MM=GM\).

Kruskal-Szekeres coordinates \(x\) and \(y\) are defined by
	\begin{subequations}\begin{align} x\,y&\ =\ \left(\frac r{2\mathcal{M}}-1\right)e^{r/2\mathcal{M}}\ , \\[3pt]
	x/y&\ =\ e^{t/2\MM}\ .
	\end{align}\end{subequations}
	
In terms of these, Eq.~\eqn{SSmetric} turns into
	\be \dd s^2\iss16\mathcal{M}^2\left(1-\frac{2\mathcal{M}}r\right)\frac{\dd x\,\dd y}{x\,y}+r^2\,\dd\W^2\iss \frac{32\MM^3}{r} \,e^{-r/2\MM}\,\dd x\dd y+r^2\,\dd\W^2\  . \ee

When we map these coordinates \(x\) and \(y\) onto the finite line segments \([-1,1]\), for instance as
	\be x=\tan(\fract{\pi}{2}u^+)\ ,\quad y=\tan (\fract{\pi}{2}u^-)\ , \ee
and we regard \(u^++u^-\) as a space-like coordinate, while \(u^+-u^-\) is time-like, we get the well-known Penrose diagram, where the lines \(x\,y=-1\) , \ \(x\ra\pm\infty\) and \(y\ra\pm\infty\) are straight lines,
while light-like geodesics with \(\dd\W=0\) run under \(45^\circ\), see Figure \ref{penrose.fig}. 

\begin{figure}[h!]\bcenter
\includegraphics[width=300pt]{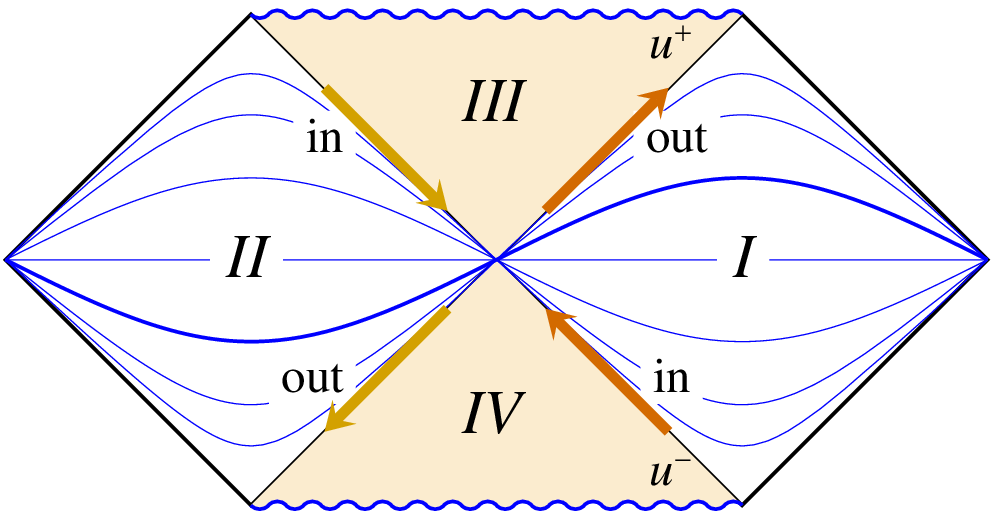}\ecenter \begin{quote}
	\begin{caption}{ \small Penrose diagram for Schwarzschild black hole, showing regions $I - IV$. Equal-time lines are shown. In the standard picture, only region \(I\) corresponds to the visible universe. In this work, region \(II\) will represent the antipodal region of the same black hole. In the classical theory, this makes no difference. Arrows show the ``firewalls", caused by the very early in-going particles (in) and the very late out-going ones (out).} \labell{penrose.fig}
	\end{caption}\end{quote}
\end{figure}

Without quantum mechanics, General Relativity clearly states what observers will see. The region labelled \(I\) in this figure represents the visible universe\fn{With apologies to the reader, as in many publications, regions \(I\) to \(IV\) are simply labelled anti-clockwise.  For us, the  labels used in Fig.~\ref{penrose.fig} are more convenient for later considerations.} It is bounded by the lines \(x\ra\infty,\ y\ra \infty\) and the future and past event horizons. All particles and fields in this universe are represented by the appropriately transformed fields in this region \(I\). The other regions are mathematical extensions of the metric, but physically, they mean rather little. Region \(III\) is seen as where an unfortunate traveller who entered the black hole may think he is going, but nothing about his adventures can be followed by observers in the outside world. Regions labelled \(II\) and \(IV\) are often considered to be even more unphysical; these are actually not there, if we may assume that the black hole was formed by collapsing matter in the (recent or distant) past. We do not need to be bothered by any of this as long as we do classical physics, and limit ourselves to epochs well after the event(s) that produced the black hole.

Also, at this point, if someone would suggest that region \(II\) corresponds to the antipodal region of the same black hole, this would not make any difference; to check such a statement, signals would have to be sent going faster than the local speed of light, and therefore, at this stage, such a statement is empty. It is not even obviously wrong.

\subsection{Hawking particles\labell{Hparticles}}
Now, switch on quantum mechanics. One important thing changes. Region \(III\) now definitely becomes important\fn{or, at least, a small region of size \(\e\) around the centre of the diagram.}. Let us assume, for the moment, that an observer going in does not encounter a `firewall'\,\cite{AMPS} due to the out-going Hawking particles. Indeed, if there were such a firewall, one would not even  be able to derive the existence of Hawking particles. It is altogether reasonable to assume that an observer passing the future event horizon, will not notice anything out of the ordinary. According to co-moving physicists, no extremely energetic particles were emitted by the collapsing body\fn{therefore, no firewalls, if we adhere to standard General Relativity, see later (Subsection~\ref{firewalls}).}, so none of such objects will be detected, if we may believe the laws of General Relativity in combination with causality.  We shall address some entanglement issues shortly.

Indeed, we do need the existence of region \(III\), smoothly seamed onto region \(I\), in order to describe the vacuum state experienced by the in-going observer. Performing the by now standard calculations by Hawking, one finds that the vacuum state as seen by the in-going observer, the Hartle-Hawking vacuum\,\cite{HH}, is seen by the distant observer as a state containing particles, roaming in a \emph{different} vacuum, the Boulware vacuum\,\cite{boulware}\fn{ We use a definition of the different possible states in the Schwarzschild metric as given by Candelas\,\cite{candelas}; in fact, as we here only look at the particles moving out, no distinction will be needed to be made between the Hartle-Hawking vacuum and the Unruh vacuum\,\cite{unruh}. }. In terms of the complete set of quantum states in the Boulware vacuum, labelled as \(|E,n\ket\), where \(E\) stands for the energy and \(n\) for any other kinds of quantum numbers, the Hartle Hawking vacuum can be written as 
	\be |\emptyset\ket_{HH}=C\sum_{n,E}|E,n\ket_I\,|E,n\ket_{II}\,e^{-\half\b_H E}\ , \eel{hawkingdistr}
where `\(HH\)' stands for `Hartle-Hawking', \(\b_H =1/kT_H=8\pi GM/\hbar c^3\) is the inverse Hawking temperature, and \(C\) stands for a constant needed to normalize the state to one.

The sign of the energies in region \(II\) has been inverted w.r.t.\ the local Lorentz boost Killing vector, because the Schwarzschild time \(t\), there, runs backwards when compared with the local time parameter \(u^+\,-\,u^-\); in our notation, the states \(|E,n\ket_{II}\) in region \(II\) are exact replicas of the states \(|E,n\ket_I\) in region \(I\).
Since the probabilities are the squares of the amplitudes of the distribution \eqn{hawkingdistr}, we recognize a thermal Boltzmann factor, \(e^{-\b_HE}\).

Recently, part of this argument has been put in doubt\,\cite{AMPS}. Arguments addressing possible entanglement between particles going in and out, in combination with no-cloning theorems, are used to claim that observers entering black holes with a long lifetime, will notice the presence of Hawking particles, and in fact be killed by the firewall formed by them. 

Even though we now claim that no firewall will arise, the tendency of very late (and very early) particles to form firewalls will play an important role in what follows, and pointing out the threat coming from these firewalls was justified.  The model we bring forward, is a natural application of General Relativity, and based on the possibility to \emph{transform away} the firewalls. The fact that, in our final results, the firewalls are to be removed, is an absolutely essential ingredient of General Relativity, and indeed we must show how this cure of the problem comes about, as we shall (Chapter \ref{micromodel}).

Yet, there is something odd. One might well accept that there cannot be a firewall on the future event horizon, but what about the past event horizon? There, we have not only a firewall, but we have the heavy bombardment of all in-going particles that participated in the collapse event that gave rise to the birth of the black hole. As fiercely as many investigators insist that there cannot be a firewall in the future, they surely accept these heavy, bombarding particles in the past. Indeed, the gravitational fields of these particles produced the black hole in the first place, and this caused the space-time metric to take a form quite different from the one drawn in Figure~\ref{penrose.fig}. Regions \(II\) and \(IV\) are usually removed from the scene. Only `eternal black holes', whatever may distinguish them from black holes formed in an `extremely distant past' are sometimes assumed to have the Penrose diagram of Fig.~\ref{penrose.fig}.

Why this difference? Whence this totally time-reversal asymmetric picture? As we argued in the Introduction, Section~\ref{intro}, we only accept a time-reversal symmetric treatment of the black hole microstates. 

There is an other problem with the picture we drew so-far, in fact, there are (at least) two problems: suppose we apply the laws of General Relativity as formulated in the books. The Boulware states, such as the ones occurring in Eq.~\eqn{hawkingdistr}, are highly divergent. In ordinary physics, such as when we calculate the Planck distribution in ordinary radiation, the Boltzmann factors at high energies provide for natural convergence, but such does not happen here. If we replace the radial coordinate by \emph{tortoise coordinates} \(\r\), to be defined as
	\be r-2GM=C\,\ex{\r}\ , \ee
then we see the in-going particles as plane waves, moving in with constant velocities (constant in terms of the Schwarzschild time variable \(t\)), and the out-going ones would move out with constant velocities. They can go all the way to \(\r\ra -\infty\), but they will never arrive there (according to an outside observer). This means that wave packets have an infinite space to occupy there. Thus, the number of Boulware states is strictly infinite. Because the gravitational potential rapidly approaches the value \(-1\), there is no energy to reduce their numbers by means of Boltzmann factors. The only way to block such an infinity would be to postulate the existence of a `brick wall'. This author investigated what such a brick wall would do\,\cite{GtHBH85}, knowing quite well that we should not really assume its existence. The brick wall had been introduced to determine what would be required to cut this infinity out.

The other difficulty is, that region \(II\) is `not really there'. It represents the `inside' of the black hole. In doing quantum mechanics, we have to average over the contributions of the quantum states there, but not being able to detect them, these states cannot be included in the final states of an evolving system. Consequently, the earliest conclusions were that black holes `turn pure quantum states into mixed states', something that a Schr\"odinger equation never does by itself.  Today, many researchers assume that the states \(|E,n\ket_{II}\) (or at least the information they contain) will eventually escape from the black hole somewhere far in region \(III\). This evidently false assumption lead to the \emph{black hole information problem} and the \emph{firewall assumption}; we have a much better scenario.

\subsection{Gravitational back reaction\labell{gravback}}

The point is that there is more. Instead of analysing the situation further -- to establish a third difficulty -- most authors now turn on their marvellous fantasy to produce one crazy model after the next. We are not ready for that yet. First, we need to contemplate the third difficulty, which is, that one cannot ignore gravitational back reaction.

Again, consider the classical theory, but now take into consideration that in-going and out-going particles all carry gravitational fields. The effects of these fields are far more bizarre than any other interactions they are involved in. The field of an in-going particle, and that of out-going ones, can be calculated precisely. All one has to do is consider the gravitational field of a particle at rest, and the way it affects the metric of the surrounding space-time, and boost that to very high velocities. The result of such a calculation is well-known: there is a \emph{dragging effect}: when an out-going particle passes in ingoing one, both particles are shifted a bit. The shift \(\d u^-\) experienced by an out-going particle, is equal to\,\cite{AichelbSexl}
	\be\d u^-_\outt=-4G\log|\tl x-\tl x\,'|\,p^-_\inn\  , \eel{rindlerdrag}
where \(G\) is Newton's constant, \(\tl x\) the transverse position of the out-going particle, and \(p^-_\inn\) is the momentum of a particle entering the hole at the transverse position \(\tl x\,'\). The distance \(|\tl x-\tl x\,'|\) is the shortest distance by which they pass (the impact parameter). The minus sign implies that the dragging is in a positive direction when that distance is small\fn{with apologies for our sign convention here; for particles coming in in region \(I\), both \(\d u^-\) and \(p^-\) are negative.}. The expression given in Eq.~\eqn{rindlerdrag} is actually what one gets in Rindler space; on the black hole horizon, one gets a slightly more complicated expression in terms of the solid angles \(\W\equiv(\tht,\vv)\) and \(\W'\), which is an angular distance. In fact, what one has here is a Green function \(f(\W,\W')\) obeying
	\be (1-\D_\W)f(\W)=\tl\d^2(\W,\W')\ ,\qquad u^-_\outt=\fract{8\pi G}{R^2}f( \W,\W')p^-_\inn\ , \eel{green}
where \(R=2GM\) is the radius of the horizon\fn{Observe that the coefficient is dimensionless because \(u^\pm\), or \(x\) and \(y\) near the horizon,  and with them also the associated momenta  \(p^\mp,\) have been defined such that they are dimensionless.}.	 The extra term 1 in this equation follows from a careful calculation of the Ricci tensor of the gravitational shock wave\,\cite{GtHDray} produced by the boosted particle(s).

Now, we must take this effect into account, because the positions \(u^+_\inn\) and \(u^-_\outt\), as well as the momenta \(p^-_\inn\) and \(p^+_\outt\) vary \emph{exponentially} with Schwarzschild time \(t\), so, for times scales of the order \(M\log M\) in natural units, this effect grows rapidly.

The most peculiar consequence of this dragging effect is, that it drags particles to and fro, across the horizons from region \(I\) and \(II\) and back, in the Penrose diagram. As soon as we try to take this effect into account, we have to abandon hopes that region \(I\) and region \(II\) can be handled independently.

Other interactions, such as the ones due to Standard Model forces, are far less drastic. They may rotate the gauge phases of wave functions, but never drag particles from one region into an other. In contrast, the gravitational force, growing exponentially in time, may generate genuine firewalls along either future or past event horizons. These firewalls look fearsome, but we shall handle them.

\subsection{Quantum dragging\labell{qudrag}}

Carrying this dragging effect over to our quantum models is relatively simple. We were accused of doing a `semiclassical' calculation, but this is not true. What we shall do is about as semiclassical as a calculation of the spectrum of a hydrogen atom using the classical electric potential fields between protons and electrons. One always takes this potential to be classical, because operators \(V(\vec x)\) are commuting operators; regarding them as creation and annihilation operators for photons in unnecessary, at least in a first attempt to calculate atomic spectra reasonably accurately. Similarly here, the Green functions \(f(\W,\W')\) in the previous section represent commuting operators in the Hilbert space of the gravitons.

Thus, we relate the \emph{positions} \(u^-(\W)\) of the out-going particles to the \emph{momenta} \(p^-(\W)\) of the in-going ones, and vice versa. Note that we can easily take the momenta of the in-going particles all to commute with one another, as well as either all momenta, or all positions of the out-going ones as being commuting, but of course, momenta do not commute with positions of the same particles; they are controlled by the familiar commutation rules.

The algebra derived from Eq.~\eqn{green} can be summarized as
	\begin{equation}\begin{aligned} 
	(1-\D_\W)\, u^-_\outt(\W)&\ =\ \fract{8\pi G}{R^2} p^-_\inn(\W)\ ,\\	
	\ [u^+_\inn(\W),\,p^-_\inn(\W')]&\ =\ i\d^2(\W,\,\W')\ ,\\
(1-\D_\W)[u^+_\inn(\W),\,u^-_\outt(\W')]&\ =\ \fract{8\pi i G}{R^2}\d^2(\W,\,\W')\ ,\\
\ [u^-_\outt(\W),\,p^+_\outt(\W')]&\ =\ i\d^2(\W,\,\W')\ ,\\
(1-\D_\W)\, u^+_\inn(\W)&\ =\ -\fract{8\pi G}{R^2} p^+_\outt(\W)\ .	
	\labell{inoutgreen}  \end{aligned}\end{equation}
	
In our previous papers we made excessive use of the fact that the dragging effects described by these equations are \emph{linear} in the momenta and the positions. This begs for an expansion in spherical harmonics. A huge advantage of spherical wave expansions is that,  expressed in terms of the spherical harmonics \(Y_{\ell m}(\W)\), the different partial waves decouple. The angular operator \(1-\D_\W\) becomes the C-number \(\ell\,^2+\ell+1\). This means that, as in the hydrogen atom, we are left with ordinary differential equations, and furthermore, near the horizon(s), we can also expand the in-going and out-going particles in terms of plane in-going and out-going waves. The resulting equations could not be simpler, and are now completely transparent. They tell us exactly what happens, for everyone to see.

Note, however, that we are not dealing with quantum wave functions here, but with operator valued waves.

\newsecl{Modelling the black hole microstates}{micromodel}

We are now in a position to formulate our procedure precisely. The question we ask is:\\
\par \emph{Using only legitimate transformations in General Relativity, can we now categorize the black hole microstates?}  And how do these evolve in time?\\

\noindent Let us ask for those quantum states that are almost stationary in Schwarzschild time \(t\). Of course, there will be one state that is completely stationary: the empty Penrose diagram of the `eternal'  black hole. For the distant observer, this state will be highly complex: as we have an `exact' Hartle Hawking vacuum\,\cite{HH} near the intersection of the past and future event horizons, this state must represent not only Hawking particles going out, but also their time-reverse, `Hawking particles' entering the black hole from infinity. Indeed, this means that energy not only flows out of the black hole, but also enters; we have a completely stationary heat bath\fn{but, our `heat bath' will not be exactly thermal, since, paradoxically, it will be in a pure state, as we shall see.}, surrounding the black hole.

In this state, Hawking particles also enter and leave section \(II\) of the Penrose diagram. We come to that.
 Now, in the Schwarzschild metric,  let us consider any quantum state of the Standard Model (or whatever theory is used near the Planck scale, as long as it allows for perturbative calculations). There will be particles travelling in, and particles going out, but we \emph{cannot allow particles whose momenta} -- in terms of the Kruskal-Szekeres coordinates \(x\) and \(y\) -- \emph{are so high that they cause a significant amount of gravitational dragging}. These would contribute to firewalls, and thus make our calculations invalid. At first sight, this demand seems to force us to abandon the condition that our states are stationary, but wait and see.
 
 We consider the most general state of light (\emph{i.e.} non energetic) particles in the Penrose diagram, with this important limit on their momenta. These are the states that will generate the microstates we want. Having a limit on the momenta will mean that a limit will be generated on the number of microstates, with one caveat: considering each partial spherical wave separately leads to ideally well-behaved states, but, at first sight, there still will be a divergence towards high values of \(\ell\) and\(m\) of the partial waves. Fortunately, these will be limited as well. The remark was made that the total number of Hawking particles ever involved with a black hole, is such that one gets not much more than one particle per Planckian surface element\,\cite{dvali}. This means that expansions beyond the values for \(\ell\) that correspond to Planckian momentum values in the transverse direction, will be unnecessary. This issue is not yet completely settled, but reinforces our conviction that this infinity can be tamed.  The effect of a cut-off in the values of \(\ell\) will be that the angular coordinates \(\W=(\tht,\vv)\) will form a lattice with Planckian dimensions.
 
Thus we assume that, exactly when positions in the transverse coordinates (the angles on the horizon) reach Planckian values, other branches of physics will have to be addressed. That happens at \(\ell\approx M\) in Planck units; all partial waves with considerably smaller values of \(\ell\) will be allowed, which is a sizeable fraction of all partial waves that may exist. Our description of the microstates will be limited to these.
 
 Now comes the good part. Consider the time evolution of the microstates we have now under consideration, and remember that every partial wave can be considered independently of all others. The in-going particles will gather alongside the past event horizon; their position operators \(u^+\) will shrink exponentially. Their momenta \(p^-\) will grow exponentially, and sooner or later, we will have to consider the gravitational dragging effect caused by them.
 
Since we are looking at a spherical wave, the calculation of the effect this gravitational dragging has on the out-going particles is easy. Everything happens at one value for \(\ell\) and \(m\) only. Their position operators \(u^-_{\ell,m}\) will grow, together with the \(p^-_{\ell,m}\) of the in-going particles. The relation between \(u^-_\outt\) and \(p^-_\inn\) stays the same, both grow exponentially. It is given by a \(2\times 2\) matrix, whose entries refer to whether the particles (more precisely, the waves) sit in region \(I\) of region\(II\). As soon as \(u^-\) and \(p^-\) become too large, we can consider the in-going particle as having become invisible (it is stuck against the past horizon), while the out-going particle has left the system. Our information retrieval process has done its job. The out-going particle carries the information of the in-going one back to the outside world. It's gone. We remove them both.
 
 In passing, note that the spherical wave describes regions on the horizon where they contribute positively and regions where they are negative; all spherical waves are both in region \(I\) and in region \(II\). The positive parts of the amplitude may describe particles going in at one side, the negative parts describe particles going in at the other side, or maybe particles removed from the first domain -- we still have to clarify what the two Penrose regions exactly mean. For the time being, note that at the horizon of a black hole there will be no positive energy theorem; energy can be positive or negative, by contrasting against the `stationary' Hartle-Hawking vacuum.
 
 We do conclude one important thing: the momenta of in- and out-going articles grow when they come close to a horizon. As soon as this happens, we replace them by the effect they have on the other articles, which are being shifted. This way, we can always limit ourselves to low momentum particles, even as time evolves: \emph{there is no firewall, either on the future or on the past horizon}  -- in-going particles threatening to produce a firewall are simply removed, while putting the information in their energy-momentum tensor in the new coordinates of the shifted out-going particles. Thus, as time evolves, we can keep using the eternal Penrose diagram of Figure~\ref{penrose.fig}.
 
 \newsecl{The antipodal identification}{antipods} If we would end our story here, there would still be a problem: what does region \(II\) stand for? Our problem is that the drag, described in the algebra \eqn{inoutgreen}, carries particles (more precisely, their spherical waves) from region \(I\) to \(II\) and back. The resulting scattering matrix is a two-by-two matrix\,\cite{GtHdiagonal}.

 Miraculously, we have two asymptotic regions with in-going and out-going particles, instead of one. This does not seem to describe the universe outside the black hole properly. Until shortly, most investigators (including this author) have been thinking that regions \(III\) and \(IV\) refer to the ``inside" of the black hole. Certainly, it seems, region \(II\) is `even further inside'. Whatever enters in region \(II\) will eventually escape, but exactly how this happens was unclear. 

In our present description, the structure of a black hole, after time separations long compared to \(M\log M\) in natural units, should be totally irrelevant, so if we would assert that what enters into region \(II\) will somehow re-appear much later, the picture would become entirely inconsistent. 

General Relativity tells us that  coordinate transformations can be used to describe regions of space and time differently. However, this only makes sense if every point that is described by one set of coordinates in one frame, should also be described by just one set in the other frame. Not two, as we have now (Note: the Schwarzschild coordinates \(r\) and \(t\) in region \(II\) are \emph{the same} \(r\) and \(t\) as in region \(I\), so, unless we do something about this, we have here a one-to-two mapping, or: the Penrose diagram seems to describe two black holes rather than one).

So what does the \emph{Einstein-Rosen bridge} (the central region of the Penrose diagram) do? Some authors suspect that region \(II\) refers to an other black hole, somewhere else in this (or another) universe. This would completely violate the most basic concept of locality, and the present author sees no way to make sense out of such suggestions. A more elegant resolution had to be searched for.  

There is an elegant solution\fn{The antipodal identification has been considered long ago, as far as we are aware, first by Sanchez and Whiting\,\cite{sanchezwhiting}. They did not mention some peculiar consequences such as the entanglement of the Hawking particles.}. It tells us that the transformation from coordinates of a distant observer, to local coordinates in the Penrose diagram, must be topologically non-trivial. Is there a problem with that? Not al all if the point mapping is one-to-one, \emph{and if it does not lead to singularities}. This is what we get if we ordain that region \(II\) of the Penrose diagram exactly describes the domain of the universe extending from the \emph{antipodes} of region \(I\). We do not have to fear that this would mean that two halves of our universe would be identical. Precisely not: the two domains \(I\) and \(II\) of the Penrose diagram are distinct. Inside the black hole, one cannot travel from one to the other, since these regions are space-like separated. Also, the points that are to be identified are always far away from one another. The closest they get is at the horizon itself, but that has a finite radius \(R=2GM\), so even there, we have no singularity. We conclude that this mapping is legitimate. Indeed, we claim that it is inevitable.  

It is also interesting. In Ref.\,\cite{GtHantipodal}, it is shown that there is a remarkable consequence for the Hawking particles. Eq.~\eqn{hawkingdistr} now implies that both the states \(|E,n\ket_I\) and \(|E,n\ket_{II}\) describe visible particles outside the horizon. This means that our state no longer requires summation over the `unseen' states to form a density matrix, but instead corresponds to a pure quantum state. 
If one sums over the other states because they are too far away for a local observer to observe (all the way to the antipodal area), then one would conclude these particles to have the usual Hawking temperature. However, the state entire is not thermal at all. There is a superb entanglement between the Hawking particles on one hemisphere of the black hole and the other. If we could do experiments with radiating black holes, this entanglement would have been detectable: if at one side of the black hole a particle emerges that happens to be strongly suppressed by a Boltzmann factor, then at the other side, the same particle will be seen to emerge with probability one -- not suppressed at all!

Indeed, the heat bath mentioned earlier, is a strangely unphysical one: antipodes in 3-space are 100\,\% entangled. In practice, this means that the stationary Hartle-Hawking state is extremely improbable in describing the universe far from the location of the black hole. All we need to conclude is, that in practice, a black hole will not be in a stationary quantum state.

And what do regions \(III\) and \(IV\) eventually stand for? Are they to be identified as well? Indeed, they are each other's antipodes, but do they truly exist? This is not so clear. When we `remove' a particle from the firewall, the regions \(III\) and/or \(IV\) will be shifted about. In a sense, these regions contain the original particles sent through the horizon. These particles now have transmitted their information onto the out-going ones. It seems that regions \(III\) and \(IV\) contain clones of the particles that we do see entering and emerging from the black hole. We claim that the question is immaterial. These regions formally exist if an in-going particle crosses its future event horizon, not bothered by a firewall because we removed that from our model, but most likely they end up being  mathematical artefacts that need not be further considered.

Indeed, we conclude that our quantum model departs quite a bit from the corresponding classical picture. This, again, shows the weirdness of quantum mechanics, but we see no contradiction in the equations anywhere.

As for the original `firewall problem', we now see that it originates from an incorrect counting of states. If we hadn't identified regions \(I\) and \(II\) with antipodal points in the black hole, every quantum state would require the existence of a clone somewhere inside the black hole (or elsewhere in the universe). Now, we have restored the one-to-one nature of the general coordinate transformation. The gravitational drag phenomenon returns information carried by particles that were about to escape to regions \(III\) or \(IV\), back into our universe, so that this quantum information is preserved.

\newsecl{Black hole information}{blinfo}

The reader may have all sorts of reasons to question our claim of the antipodal identification. At least the present author did not pay much attention to the publications by Sanchez and Whiting when they appeared. Our point is now that the spherical wave expansion forces this interpretation on us. Having the two regions, \(I\) and \(II\), with the gravitational dragging mixing them up, yields a \(2\times 2\) matrix relating in- to out particles. One clearly sees that unitarity is only restored if the out-going particles in region \(II\) follow their way in the negative time, that is, negative \((u^+\,-\,u^-)\)  direction. Thus, we have to identify region \(II\) to some other part of the same black hole. keeping the same Schwarzschild time coordinate, which, in the Penrose diagram, flips direction. It sounds too odd to be true, but it is the only picture that works. Note, that we were forced to solve \(2\,\times\, 2\) problems: all four regions, \(I\) -- \(IV\) needed to be understood.

Did we solve the black hole information problem? Do we now have a Schr\"odinger equation to describe the evolution of the black hole microstates, with a hermitean Hamiltonian and a unitary evolution operator? We claim to have come close to that, but there is still lots of work to do. The primary problem is as follows.

We postulated that a particle ending its useful life on a horizon, passes on its information content to the out-going particles by the gravitational dragging effect. In the early days, we used functional integrals to describe this mechanism. The in- and out-going particles were found to contribute just as vertex insertions, on the horizon that resembles the world sheet of a string. Just as string theory produces unitary amplitudes out of these string world sheet expressions, our microstates should also end up as evolving in a unitary way.

However, this is a dangerously big step to take; we promised small steps only. Therefore, we would like first to see more explicitly how it can be that the Standard Model particles, with all their quantum numbers, can leave their imprint on the out-going particles exclusively through their momentum distribution, which we describe in terms of spherical partial waves. 

This part of our proposed mechanism has not yet been described well: how do we transform the information of all fields of Standard Model particles, as well as the perturbative gravitons, in terms of momentum distribution functions only?

It is important to distinguish first quantization from second quantization. In a second-quantized theory, such as the Standard Model, the generic quantum state can have variable numbers (including zero) of particles at any point of its coordinate system. However, our gravitational drag effect cannot distinguish many particle states from single particles states; every partial wave with quantum numbers \(\ell\) and \(m\) leaves exactly one imprint. Consequently, in the longitudinal space, spanned by the coordinates \(u^+\) and \(u^-\), each stretching over both regions \(I\) and \(II\), we have exactly one amplitude for each value of \(\ell\) and \(m\). At any given  \((\ell,\,m)\), the coordinates \(u^+_{\ell,m}\) and \(u^-_{\ell,m}\) do not commute, so we need to specify one of them to describe an orthonormal set of elements of our Hilbert space. It is this Hilbert space that needs to be mapped one-to-one on the elements of the Standard Model Hilbert space.

\begin{figure}\begin{box}2 \begin{center}		{\small
\(\qqquad\  A\hfill B\hfill C\hfill D\hfill E\ \ \qquad\quad\ \)\\[8pt]
\def\pijl#1{\begin{matrix}\ \quad\ra\ \quad \\ #1 \end{matrix}}
\(\begin{matrix}\hbox{SM}\\ \hbox{particles}\\ +\\ \hbox{gravitons}\end{matrix} \pijl{(1)}
\begin{matrix}  p^-(\tht,\vv)\\ \hbox{on future}\\ \hbox{event}\\ \hbox{horizon,}\\ \hbox{firewall on}\\ \hbox{past event}\\ \hbox{horizon}
\end{matrix} \pijl{(2)}  \begin{matrix}
u^-(\tht,\vv)\\ \hbox{position of}\\ \hbox{particles}\\ \hbox{on past}\\ \hbox{event}\\ \hbox{horizon} \end{matrix}
\pijl{(3)}  \begin{matrix}
p^+(\tht,\vv)\\ \hbox{momentum}\\ \hbox{of particles}\\ \hbox{on past}\\ \hbox{event}\\ \hbox{horizon} \end{matrix}
 \pijl{(4)} \begin{matrix}\hbox{SM}\\ \hbox{particles}\\
 +\\ \hbox{gravitons} \end{matrix} \)  } \end{center}\end{box}
\begin{caption} {Steps in the evolution of microstates.\labell{steps.fig}} \end{caption} \begin{quotation} \noindent
{\small (1):  \(p^-(\tht,\vv)\) is calculated from the energy-momentum tensor of the SM particles. \\
(2):  Effect of the gravitational drag \\
 (3): Fourier transformation to momentum space \\
 (4): invert step (1). } \end{quotation}
\end{figure}

Figure \ref{steps.fig} now shows how the evolution proceeds through various steps. All of these steps are fairly straightforward now, with the exception of the last one. To recover the quantum states of the particles from their energy momentum distribution at the past event horizon is not obviously possible in the generic case. However, we can imagine how an experimental physicist may recognize elementary particles from their tracks; similarly, it may well be possible that the Standard Model particles, and even the presence of gravitons, can be recognized from the patterns in the momentum distribution. In principle, this is quite well possible; all we need to ordain is that the dimensionality of both Hilbert spaces be the same (in finite black holes, it is expected that these spaces have finite dimensionalities).

In practice, one may imagine things to go as follows.
Suppose we had a detector with Planckian size resolution, but it only detects momentum distributions. An \(N\)-particle state should be easy to recognize: there will be \(N\) peaks in the distribution \(p^+(\tht,\vv)\) of this state. Then, one might assume that, at the Planck scale, there is a basic spinless particle as the most elementary building block of matter. Starting from all \(N\)-particle states of this particle, one can build all states of the composites. 

Of course, the real situation might be quite a bit more complicated, but in principle it is not hopeless to suspect that this can be used to restore unitarity in the evolution of these states.

In more general terms, we simply have the situation that there are two high-dimensional Hilbert spaces, one consisting of all Fock space elements of the Standard Model particles (plus gravitons), and the other consisting of all partial waves with quantum numbers \(\ell\) and \(m\), each with their specific amplitudes. There should exist a precise, unitary transformation connecting these two spaces.

In Ref.\,\cite{GtHgravshift}, it was found that this unitary transformation appears to be very similar to a closed string amplitude, where the black hole horizon acts as the string world sheet. This raises the interesting possibility that string theory might provide the answer as to how the unitary transformation we are searching for, can be constructed in practice.

\newsecl{Miscellaneous remarks}{misc}
\subsection{Black Hole Complementarity\labell{compl}} The way we previously thought of black hole complementarity consisted of two ingredients:
\bi{(1)}Let \(\HH_1\) be the complete set of states in a Hilbert space containing a black hole together with all elementary particles surrounding it,  as observed by a distant observer, and \(\HH_2\) the complete set of quantum states observed by a local observer residing close to the intersection region of future and past event horizons, or  falling into a black hole. Then there should be a unitary mapping \(U\) connecting these two Hilbert spaces, similar to the mapping of field space and Fock space for Standard Model particles.
\itm{(2)}Consider the causal ordering of two point-events \(x_1\) and \(x_2\) in space-time, as seen by these two observers. \(x_1\) could be in the future light cone of \(x_2\), or \(x_1\) could be in the past light cone, or \(x_1\) and \(x_2\) could be space-like separated. The two descriptions for the time evolution as defined in point (1), should both make use of space-time points, and the causal relationships should be agreed upon by both observers. \ei
The reader may have noticed that point (1) is completely valid in our description, but we depart from point (2). In region \(II\) the causal order has been inverted. This is a curious and unexpected situation, but the explicit calculations of Refs.\,\cite{GtHantipodal} and \cite{GtHdiagonal}, to which we refer now, demand it this way.

\subsection{Background dependence\labell{backgr}}
Often, big distinctions are made as to whether a theory is \emph{background metric dependent} of \emph{independent}. Eventually, we wish to understand the subtle relationship between material particles and the space-time metric fluctuations caused by them. Most theories, such as perturbative quantum gravity and even string theory, require a fixed background for a proper formulation of their dynamical laws. The same is true in the present theory. However, we do include a non-perturbative treatment of the gravitational shift effects. These are included in the dynamics, so, at least the non-trivial consequences of this kind of space-time curvature is handled \emph{non-perturbatively} here. For instance, the in- and out-going particles cause shifts that may grow exponentially in time, eventually causing particles to disappear into asymptotic infinity. This gravitational effect is therefore topologically non-trivial, as is the topological phenomenon of antipodal identification. The new ingredient of our theory is that only in a limited time segment (regarding the Schwarzschild time coordinate \(t\)), we keep the background metric fixed, but at greater time separations, the gravitational shifts become essential.

Note that the only physical region that we need for our description are regions \(I\) and \(II\) of the Penrose diagram, Fig.~\ref{penrose.fig}, plus a region that one could call \((\e)\), the immediate neighbourhood of the intersection between past and future event horizons.

The inside of the black hole has essentially disappeared. All that is left, is the horizon, a perfect explanation as to why it is, that the entropy is proportional to the horizon area and not the bulk volume. There is no bulk volume.

\subsection{Firewalls\labell{firewalls}}
The apparent emergence of strong entanglement between Hawking radiation emitted at different epochs of a black hole's existence, has caused for some stir in the literature\,\cite{AMPS}. If Hawking radiation would emerge in a pure state, as, it seems, should follow from the scattering matrix ansatz, then this would be a big departure from its description in terms of a Hartle-Hawking state. The pure Hartle-Hawking state would lead to more entanglement than these researchers can handle, so that the fear arises for quantum cloning, which should never be a feature of the states contemplated by any single observer.

We insist that our way of describing the general behaviour of the states in the neighbourhood of a black hole, resolves this problem. The Hawking particles are not precisely thermal, since there is a strong entanglement between the particles emitted at opposite hemispheres of the horizon. Indeed, according to Eq.~\eqn{hawkingdistr}, the states turn out to be pure quantum states, when the two hemispheres are combined. This does not violate Einstein's equivalence principe because the points at opposite sides of the horizon are always far apart, so that no singularity emerges, and \emph{locally}, we always have pure coordinate transformations.

When handled this way, a black hole behaves in no way differently from any `ordinary' quantum object obeying standard Schr\"odinger equations. The fact that physical information does seem to travel `instantaneously' from a point on the horizon to its antipode, also does not violate any basic principle because the time this information takes as seen by a distant observer, approaches infinity, since it happens in terms of the tortoise coordinates. Note that, formally, in a background that is not Lorentz invariant, velocities greater than light are not directly in conflict with special relativity.

\subsection{Non commutative geometry\labell{noncomm}}
The gravitational drag effect is described by the equations \eqn{inoutgreen}. In terms of the partial wave expansion coefficients, the amplitudes of these waves are written as 
	\be 	u^\pm(\tht,\vv)&=&\sum_{\ell,m}u^\pm_{\ell m}Y_{\ell m}(\tht,\vv)\ , \ee
so that
	\be u^\pm_{\ell m}&=&\mp\frac{8\pi G/R^2}{\ell^2+\ell+1}p^\pm_{\ell m} \ ;\\[5pt]
	\  [u^+_{\ell m}\,,\,u^-_{\ell\,'\, m\,'}] &=&\frac{8\pi G\,i}{R^2}\, \frac{\d_{\ell\, \ell\,'}\,\d_{m\,m\,'}} {\ell^2+\ell+1}\ . \eel{commuu}

This last equation reminds us of non-commutative coordinates, but beware, \(u_{\ell m}^\pm\) are not just coordinates but rather amplitudes of spherical waves, describing more than just a single particle. So, while often non-commutative coordinates are suggested to accommodate for some kind of space-time grid, they here come as derived from well-established physical arguments, and the resemblance is not more than superficial. In particular, the \(\ell\)-dependence must be noted.

\subsection{Standard Model forces\labell{SMforces}}
	When we talk of Standard Model particles, as in Fig.~\ref{steps.fig}, what we really mean is whatever particle scheme is opportune at distance scales not too far out from the Planck scale. It may well be that we have some sort of unified field theory here, such as \(SU(5)\) or \(SO(10)\). At this stage of our research the details are unimportant; this is why we should put ``Standard Model" in quotation marks. One could hope that, in due time, something quite non-trivial can be deduced to put constraints on the particle model that can be employed here, but we are still a far way off from being able to do that.
	
	The out-going particles will cross the in-going ones, and we computed the way these beams interact gravitationally. In Ref.~\cite{GtHSmatrix}, it is suggested that these forces also receive contributions from the ``Standard Model" interactions, and some preliminary computations were done\,\cite{GtHSmatrix}. Instead of trying to compute the details of steps (1) and (4) in Figure~\ref{steps.fig}, we could try to accommodate for the details of the ``Standard Model" forces by also expanding the gauge charges in partial waves and add their effects to our algebra. This has not yet been done; what we expect is that the ``Standard Model" fields will add to our algebra \eqn{inoutgreen} a `spectator algebra' that does not affect the dynamics much, but may be useful in linking our spherical waves to the spectrum of Standard Model particle states.
	
\subsection{Black hole entropy\labell{entropy}}
	Finally, there still is an important problem to be addressed: since our sums over the angular quantum numbers \(\ell\) still diverge\,\cite{GtHdiagonal}, we do not get a quantitatively accurate match with the area law of the Hawking entropy of the black hole. This is peculiar, because the same mathematical transformations are being applied here. This probably has to do with our present inability to link our microstates directly with the ``Standard Model" states (Section \ref{blinfo}). An improved formalism for counting our states should be found to cure this difficulty.\\[10pt]

\noindent {\large \textbf{Acknowledgement}}\\[10pt]
The author had useful discussions with  A.~Ashtekar, G.\ Dvali, P.\ Betzios and N.\ Gaddam.

\end{document}